\documentclass[a4paper,12pt]{article}

\usepackage{amssymb}
\usepackage{amsfonts}
\usepackage{amsmath}
\usepackage{amsthm}
\usepackage{hyperref,natbib}
\usepackage[utf8]{inputenc}

\newtheorem{theorem}{Theorem}[section]

\newtheorem{corollary}[theorem]{Corollary}

\newtheorem{definition}[theorem]{Definition}

\newtheorem{lemma}[theorem]{Lemma}

\newtheorem{proposition}[theorem]{Proposition}
\theoremstyle{remark}
\newtheorem{example}[theorem]{Example}
\newtheorem{remark}[theorem]{Remark}

\title{Young's axiomatization of the Shapley value - a new proof\thanks{I thank Ferenc Forgó, Anna Khmelnitskaya, Zsófia Széna and William Thomson for their suggestions and remarks. Financial support by the Hungarian Scientific Research Fund (OTKA) and the János Bolyai Research Scholarship of the Hungarian Academy of Sciences is also gratefully acknowledged.}}
\author{Miklós Pintér \\
Corvinus University of Budapest\thanks{Department of Mathematics, Corvinus University of Budapest, 1093 Hungary, Budapest, Fővám tér 13-15., miklos.pinter@uni-corvinus.hu}}

\begin{document}

\maketitle

\begin{abstract}
We consider \cite{Young1985}'s characterization of the Shapley value, and give a new proof of this axiomatization. Moreover, as applications of the new proof, we show that \cite{Young1985}'s axiomatization of the Shapley value works on various well-known subclasses of $TU$ games.
\end{abstract}

\section{Introduction}

In this paper we consider one of the well-known characterizations of the Shapley value \citep{Shapley1953}: \cite{Young1985}'s axiomatization. The Shapley value is probably the most popular one-point solution (value) of transferable utility ($TU$) cooperative games (henceforth games). It is applied in various fields ranging from medicine to statistics, from engineering to accounting etc. Therefore a solid characterization could well serve, among others, applications by helping in understanding its very nature.

\cite{Young1985} axiomatizes the Shapley value with three axioms: \textit{Efficiency} (\textit{Pa\-reto optimality} $PO$), \textit{Symmetry} or \textit{Equal Treatment Property} $ETP$ (although Symmetry and $ETP$ are different axioms, they are equivalent for $TU$ games), and \textit{Strong Monotonicity}. \cite{Moulin1988} suggests an alternative proof for \cite{Young1985}'s result in the three player setting. Both \cite{Young1985} and \cite{Moulin1988} consider the whole class of $TU$ games, however, \cite{Young1985} also shows that this characterization is valid on other classes of games, he specifies the class of superadditive games.

With respect to other subclasses on which \cite{Young1985}'s axiomatization works we have to mention three other papers. \cite{Neyman1989} shows that a solution defined on the additive group generated by a game is Efficient ($PO$), Symmetric (or $ETP$) and Strongly Monotonic if and only if it is the Shapley value. \cite{Khmelnitskaya2003} proves that \cite{Young1985}'s axiomatization works on the class of non-negative constant-sum games with non-zero worth of grand coalition and on the (entire) class of constant-sum games. Furthermore, \cite{Mlodak2003} applies the  same method as that \cite{Khmelnitskaya2003} does to characterize the Shapley value á la \cite{Young1985} on the class of non-negative bilateral games.

It is well-known that the validity of an axiomatization can vary from subclass to subclass, e.g. \cite{Shapley1953}'s axiomatization of the Shapley value is valid on the class of monotone games but not valid on the class of strictly monotone games. Therefore, we must consider each subclass of games one by one. 

The main motivation of this paper is methodological. We give a new proof of \cite{Young1985}'s axiomatization of the Shapley value; by this result there are three different methods for checking the validity of \cite{Young1985}'s axiomatization of the Shapley value (on subclasses of games): \cite{Young1985}'s, \cite{Moulin1988}'s and our. We emphasize these three methods are not comparable, for each of them there are cases where the one works and the others do not, vice versa, and naturally there cases where all work and where non of them works.  

The setup of the paper is as follows. In Section \ref{sec1} we introduce the terminology used throughout the paper. Section \ref{Y} discusses our main result.

\section{Preliminaries}\label{sec1}

Notation: let $| N |$ and $2^N$ denote the cardinality of set $N$ and the set of all subsets of $N$ respectively. Moreover, $A \subset B$ means $A \subseteq B$, but $A \neq B$. and we also use $|a|$ for the absolute value of real number $a$. Finally, $\sum_{i \in \emptyset} x_i = 0$, that is, the empty sum is zero.

Let $N \neq \emptyset$, $| N | < \infty$ and $v : 2^N \rightarrow \mathbb{R}$ be a function such that $v(\emptyset) = 0$. Then $N$ and $v$ are called set of players and \textit{transferable utility cooperative game} (henceforth game) respectively. The class of games with player set $N$ is denoted by $\mathcal{G}^N$.

Let $v \in \mathcal{G}^N$, $i \in  N$, and for each $S \subseteq N$: let $v'_i (S) = v(S \cup \{i\}) - v(S)$. Then $v'_i$ is called Player $i$'s \textit{marginal contribution function} in game $v$. In other words, $v_i'(S)$ is Player $i$'s marginal contribution to coalition $S$ in game $v$.

In this paper, along with $\mathcal{G}^N$, we consider also subclasses of games defined below. A game $v \in \mathcal{G}^N$ is

\begin{itemize}
\item \textit{essential}, if $v(N) > \sum \limits_{i \in N} v(\{i\})$,

\item \textit{convex}, if for each $S,T \subseteq N$: $v(S) + v(T) \leq v(S \cup T) + v (S \cap T)$,

\item \textit{strictly convex}, if for each $S,T \subseteq N$, $S \nsubseteq T$, $T \nsubseteq S$: $v(S) + v(T)$ $< v(S \cup T) + v (S \cap T)$,

\item \textit{superadditive}, if for each $S,T \subseteq N$, $S \cap T = \emptyset$: $v(S) + v(T) \leq v(S \cup T)$,

\item \textit{strictly superadditive}, if for each $S,T \subseteq N$, $S,T \neq \emptyset$, $S  \cap T = \emptyset$: $v(S) + v(T) < v(S \cup T)$,

\item \textit{weakly superadditive}, if for each $S \subseteq N$, $i \in N \setminus S$: $v(S) + v(\{i\})$ $\leq v(S \cup \{i\})$,

\item \textit{strictly weakly superadditive}, if for each $S \subseteq N$, $S \neq \emptyset$, $i \in N \setminus S$: $v(S) + v(\{i\}) < v(S \cup \{i\})$,

\item \textit{monotone}, if for each $S,T \subseteq N$, $S \subseteq T$: $v(S) \leq v(T)$,

\item \textit{strictly monotone}, if for each $S,T \subseteq N$, $S \subset T$: $v(S) < v(T)$,

\item \textit{additive}, if for each $S,T \subseteq N$, $S  \cap T = \emptyset$: $v(S) + v(T) = v(S \cup T)$,

\item \textit{weakly subadditive}, if for each $S \subseteq N$, $i \in N \setminus S$: $v(S) + v(\{i\})$ $\geq v(S \cup \{i\})$,

\item \textit{strictly weakly subadditive}, if for each $S \subseteq N$, $S \neq \emptyset$, $i \in N \setminus S$: $v(S) + v(\{i\}) > v(S \cup \{i\})$,

\item \textit{subadditive}, if for each $S,T \subseteq N$, $S \cap T = \emptyset$: $v(S) + v(T) \geq v(S \cup T)$,

\item \textit{strictly subadditive}, if for each $S,T \subseteq N$, $S,T \neq \emptyset$, $S  \cap T = \emptyset$: $v(S) + v(T) > v(S \cup T)$,

\item \textit{concave}, if for each $S,T \subseteq N$: $v(S) + v(T) \geq v(S \cup T) + v (S \cap T)$,

\item \textit{strictly concave}, if for each $S,T \subseteq N$, $S \nsubseteq T$, $T \nsubseteq S$: $v(S) + v(T)$ $> v(S \cup T) + v (S \cap T)$.
\end{itemize}

For the definition of essential games see e.g. \cite{NeumannMorgenstern1953}, and
for other types of games see e.g. \cite{PelegSudholter2003}.

The following alternative definitions of (strictly) convex and (strictly) concave games are well known:

\begin{equation}\label{lemma50}
\begin{array}{l}
\text{Game $v \in \mathcal{G}^N$ is (strictly) convex, if for each $i \in N$, $T,Z \subseteq N \setminus \{i\}$} \\
\text{such that $Z \subset T$: $v_i' (Z) \leq v_i' (T)$ ($v_i' (Z) < v_i' (T)$),} \\
\text{and $v \in \mathcal{G}^N$ is (strictly) concave, if for each $i \in N$, $T,Z \subseteq N \setminus \{i\}$} \\
\text{such that $Z \subset T$: $v_i' (Z) \geq v_i' (T)$ ($v_i' (Z) > v_i' (T)$).} \\
\end{array}
\end{equation}

The \textit{dual} of game $v \in \mathcal{G}^N$ is the game $\bar{v} \in  \mathcal{G}^N$ such that for each $S \subseteq N$: $\bar{v} (S) = v(N) - v(N \setminus S)$.

For any game $v \in \mathcal{G}^N$, players $i,j \in N$ are \textit{equivalent} (\textit{symmetric}), $i \sim^v j$, if for each $S \subseteq N$ such that $i,j \notin S$: $v'_i (S) = v'_j (S)$. It is easy to verify that for any game $v \in \mathcal{G}^N$ $\sim^v$ is a binary equivalence relation on $N \times N$.

Furthermore, if $S \subseteq N$ is such that for all $i,j \in S$: $i \sim^v j$, then we say that $S$ is an \textit{equivalence set} in game $v$.

Next we summarize some important properties of dual games. For any game $v \in \mathcal{G}^N$:

\begin{equation}\label{dualka}
\begin{array}{l}
\text{If $i \sim^v j$, then $i \sim^{\bar{v}} j$.} \\ 
\text{If $w_i' = v_i'$, then $\bar{w}_i' = \bar{v}_i'$.} \\
\text{The dual of a (strictly) convex game is a (strictly) concave game.} \\
\text{The dual of a (strictly) concave game is a (strictly) convex game.}
\end{array}
\end{equation}

Function $\psi : A \rightarrow \mathbb{R}^N$, defined on set $A \subseteq \mathcal{G}^N$, is a \textit{solution} on $A$. Throughout the paper we consider single-valued solutions (values).

For any game $v \in \mathcal{G}^N$ the \textit{Shapley solution} $\phi$ is given by

\begin{equation*}
\phi_i (v) = \sum \limits_{S \subseteq N \setminus \{i\}} v'_i  (S) \dfrac{|S| ! (|N \setminus S| - 1)!}{|N| !}\ , \mspace{20mu} i \in N,
\end{equation*}

\noindent where $\phi_i (v)$ is also called Player $i$'s \textit{Shapley value} \cite{Shapley1953}.

The solution $\psi$ on class of games $A \subseteq \mathcal{G}^N$ satisfies

\begin{itemize}
\item \textit{Pareto optimality} ($PO$), if for each $v \in A$: $\sum \limits_{i \in N} \psi_i (v) =  v(N)$,

\item \textit{Equal Treatment Property} ($ETP$), if for all $v \in A$, $i,j \in N$: $i \sim^v j$ implies $\psi_i(v) = \psi_j(v)$,

\item \textit{Marginality} ($M$), if for all $v,w \in A$, $i \in N$: $v'_i = w'_i$ implies $\psi_i (v) = \psi_i (w)$.
\end{itemize}

\begin{remark}\label{rem1}
Notice that the Shapley solution is (completely) determined by the players' marginal contribution functions. Therefore for any solution $\psi$ meeting axiom $M$: if $\psi_i (v) = \phi_i (v)$ and $v'_i = w'_i$, then $\psi_i (w) = \phi_i (w)$.
\end{remark}

It is well known and not difficult to check that the Shapley solution meets axioms $PO$, $ETP$ and $M$.

\section{The main result}\label{Y}

In this section we present our main result. The following example illustrates the idea behind our result (Theorem \ref{tetel1}). This example shows that we can construct chains of games such that in any chain the elements are connected by axiom $M$ and in the terminal games all players are equivalent.    

\begin{example}\label{pl1}
Let $N = \{1,2,3\}$ and  $v = (0,0,0,3,1,2,3) \in \mathcal{G}^N$, where $v = (v(\{1\}),v(\{2\}),v(\{3\}),v(\{1,2\}),v(\{1,3\}),v(\{2,3\}),v(N))$. Then $v$ is a superadditive but not convex game, and $1 \nsim^v 2$, $1 \nsim^v 3$, $2 \nsim^v 3$.

Furthermore, let $\psi$ be a $PO$, $ETP$ and $M$ solution on $\mathcal{G}^N$. We show that $\psi_2 (v) = \phi_2 (v)$.

Take Player $1$ as a singleton equivalence set in game $v$ and choose Player $2$. Then there is a game $w = (0,0,0,3,2,2,4)$ such that $w_2' = v_2'$ and $1 \sim^w 2$ (it is clear that $w$ is not the only game in which players $1$ and $2$ are equivalent and $w'_2 = v_2'$).

Next take equivalence set $\{1,2\}$ in game $w$ and choose Player $3$. Then there is a game $z = (0,0,0,2,2,2,3)$ such that $z_3' = w_3'$ and $1 \sim^z 2 \sim^z 3$.

Then axioms $PO$ and $ETP$ imply that $\psi (z) = \phi (z)$. Moreover, by axiom $M$, $\psi_3 (w) = \phi_3 (w)$. Since $\psi$ is $PO$ and $ETP$, $1 \sim^w 2$, therefore $\psi (w) = \phi (w)$.

By applying axiom $M$ again, we get $\psi_2 (v) = \phi_2 (v)$.
\end{example}

From Example \ref{pl1} it is clear that we can deduce $\psi_i = \phi_i$ for any $i$ (player). In other words, we can show that $\psi (v) = \phi (v)$. All we need is that $\psi$ must be defined on the paths from $v$ to $z$ (different $w$ and $z$ for different $i$).

The next notion is an important ingredient of our main theorem.

\begin{definition}\label{zartsag}
Class $A \subseteq \mathcal{G}^N$ is \textit{$M$-closed}, if for each game $v \in A$, equivalence set in $v$ $S \subseteq N$, and Player $k \in N \setminus S$ there exists $w \in A$ such that $S \cup \{k\}$ is an equivalence set in $w$ and $w_k' = v_k'$.
\end{definition}

\begin{remark}\label{rem2}
Notice that from \eqref{dualka}, if $A \subseteq \mathcal{G}^N$ is an $M$-closed class of games, then $\bar{A} = \{ \bar{v} \in \mathcal{G}^N : v \in A\}$ is also $M$-closed. 
\end{remark}

The following theorem is our main result.

\begin{theorem}\label{tetel1}
Let class $A \subseteq \mathcal{G}^N$ be $M$-closed. Then solution $\psi$ defined on class $A$, satisfies axioms $PO$, $ETP$ and $M$, if and only if $\psi = \phi$, that is, if and only if, it is the Shapley solution.
\end{theorem}

\begin{proof}[Proof of Theorem \ref{tetel1}]
If : It is well-known.

Only if: Class $A$ is $M$-closed, therefore there exists $z(1) \in A$ such that $z(1)_{i_2}' = z_{i_2}'$ and $\{i_1,i_2\}$ is an equivalence set in $z(1)$. Let $i_3 \in N \setminus \{i_1,i_2\}$.

Class $A$ is $M$-closed therefore there exists $z(2) \in A$ such that $z(2)_{i_3}' = z(1)_{i_3}'$ and $\{i_1,i_2,i_3\}$ is an equivalence set in $z(2)$. Let $i_4 \in N \setminus \{i_1,i_2,i_3\}$.

$\vdots$

Class $A$ is $M$-closed therefore there exists $z(n-1) \in A$ such that $z(n-1)_{i_n}'$ $= z(n-2)_{i_n}'$ and $\{i_1,i_2,\ldots,i_n\}$ $(= N)$ is an equivalence set in $z(n-1)$.

By axioms $PO$ and $ETP$, $\psi(z(n-1)) = \phi (z(n-1))$, since all the players $i_1,i_2,\ldots, i_n$ are symmetric in $z(n-1)$.

Since solution $\psi$ meets axiom $M$, and by construction $z(n-1)'_{i_n}$ $= z(n-2)'_{i_n}$, it follows that (see Remark \ref{rem1}) $\psi_{i_n} (z(n-2)) = \phi_{i_n} (z(n-1))$. Next, all the players $i_1,i_2,\ldots,i_{n-1}$ are symmetric in game $z(n-2)$, whence by axioms $ETP$ and $PO$ we get $\psi(z(n-2)) = \phi (z(n-2))$.

By applying a similar reasoning as above, and since $i_{n-2}$ was arbitrarily chosen, we get $\psi(z(n-3)) = \phi (z(n-3))$.

$\vdots$

Since $i_2 \in N \setminus \{i_1\}$ was arbitrarily chosen, $\psi$ is defined on $A$, and satisfies axioms $PO$ and $M$, we get $\psi (z) = \phi (z)$.

Finally, $i_1$ was arbitrarily chosen, therefore $\psi (v) = \phi (v)$.
\end{proof}

Next, we show that the above theorem implements Young's \cite{Young1985} result.

\begin{theorem}[\cite{Young1985}]\label{tetel2}
Solution $\psi$ on $\mathcal{G}^N$ satisfies axioms $PO$, $ETP$ and $M$, if and only if $ \psi = \phi$, that is, if and only if, it is the Shapley solution.
\end{theorem}

To prove Theorem \ref{tetel2} it is enough to show that $\mathcal{G}^N$ is $M$-closed.

\begin{proposition}\label{lemma3}
The class fo games $\mathcal{G}^N$ is $M$-closed.
\end{proposition}

First we prove the following lemma.

\begin{lemma}\label{lemma1}
Let $v \in \mathcal{G}^N$. Then $S \subseteq N$ is an equivalence set in game $v$, if and only if for all $T,Z \subseteq N$ such that $T \setminus S = Z \setminus S$ and $|T|  = |Z|$: $v (T) = v (Z)$.
\end{lemma}

\begin{proof}
If: It is left for the reader.

Only if: W.l.o.g. we can assume that $T \setminus Z \neq \emptyset$ (if $T \setminus Z = \emptyset$, then the proof ends), and let $T \setminus Z = \{l_1,\ldots,l_m\}$ and $Z \setminus T = \{q_1,\ldots,q_m\}$. Here, $T \setminus Z \subseteq S$ and $Z \setminus T \subseteq S$, $S$ is an equivalence set in $v$, hence for each player $i$, $1 \leq i \leq m$:

\begin{equation*}
\begin{split}
v((T \cap Z) \cup \{l_1,\ldots,l_i\}) \\
= v((T \cap Z) \cup \{l_1,\ldots,l_{i-1}\}) + v_{l_i}' ((T \cap Z) \cup \{l_1,\ldots,l_{i-1}\}) \\ 
= v((T \cap Z) \cup \{q_1,\ldots,q_{i-1}\}) + v_{q_i}' ((T \cap Z) \cup \{q_1,\ldots,q_{i-1}\}) \\
= v((T \cap Z) \cup \{q_1,\ldots,q_i\})
\end{split}
\end{equation*}

\noindent Therefore $v(T) =  v (Z)$.
\end{proof}

Next, we consider a direct corollary of Lemma \ref{lemma1}.

\begin{corollary}\label{lemma2}
Let $v \in \mathcal{G}^N$, $S \subseteq N$ be an equivalence set in $v$, and $k \in N \setminus S$. Then for all $T,Z \subseteq N$ such that $T \setminus S = Z \setminus S$ and $|T| = |Z|$: $v_k' (T) = v_k' (Z)$.
\end{corollary}

\begin{proof}[Proof of Proposition \ref{lemma3}]
Let $v \in \mathcal{G}^N$ be such that $S \subset N$ is an equivalence set in $v$, and $k \in N \setminus S$.

Let game $w \in \mathcal{G}^N$ be defined as follows for each coalition $T \subseteq N$: if $T \cap (S \cup \{k\}) = \emptyset$, then let $w(T)$ be arbitrarily defined such that $w (\emptyset) = 0$. In the other cases ($T \cap (S \cup \{k\}) \neq \emptyset$), let 

\begin{equation}\label{eq1}
w (T) = w(T \setminus (S \cup \{k\})) + \sum \limits_{i=1}^m v_{k}' ((T \setminus (S \cup \{k\})) \cup \{l_1,\ldots,l_{i-1}\}) \ ,
\end{equation}

\noindent where $m = |(S \cup \{k\}) \cap T|$, and $l_i \in S \cap T$, $i=1,\ldots,m-1$. 

Notice that from Corollary \ref{lemma2} $\sum \limits_{i=1}^m v_{k}' ((T \setminus (S \cup \{k\})) \cup \{l_1,\ldots,l_{i-1}\})$ does not depend on the ordering of the elements of $S \cap T$, that is, $w$ is well-defined.

It is easy to verify that $w_k' = v_k'$. Furthermore, from Lemma \ref{lemma1} $S \cup \{k\}$ is an equivalence set in $w$.
\end{proof}

\begin{proof}[Proof of Theorem \ref{tetel2}]
See Theorem \ref{tetel1} and Proposition \ref{lemma3}.
\end{proof}

Next, we show that \cite{Young1985}'s axiomatization is also valid on some considered subclasses of games.

\begin{theorem}\label{tetel3}
Solution $\psi$ defined on the class of either (strictly) convex, (strictly) weakly superadditive, (strictly) mon\-otone, additive, (strict\-ly) weakly subadditive or (strictly) concave games satisfies axioms $PO$, $ETP$ and $M$, if and only if $ \psi = \phi$, that is, if and only if it is the Shapley solution.
\end{theorem}

\begin{proof}
We show that all considered subclasses of games are $M$-closed.

Let $v \in \mathcal{G}^N$ be such that $S \subset N$ is an equivalence set in $v$, and $k \in N \setminus S$. The proof of Lemma \ref{lemma3} shows that there exists $w \in \mathcal{G}^N$ such that $S \cup \{k\}$ is an equivalence set in $w$, $w_k' = v_k'$, and for each coalition $T$ such that $T \cap (S \cup \{k\})= \emptyset$, $T \neq \emptyset$: $w(T)$ is arbitrarily defined. Therefore, the only thing we have to do is to show that we can give values to these coalitions such that $w$ be in the considered class of games.

\bigskip

(I) The class of additive games: It is well known that game $z \in \mathcal{G}^N$ is additive, if and only if for each Player $i \in N$ there exists $c_i \in \mathbb{R}$ such that for each coalition $T \subseteq N \setminus \{i\}$: $z_i' (T) = c_i$.

Let $c^\ast = v_k' (\emptyset)$, and for for each $T \subseteq N$ let

\begin{equation}\label{eqkell}
w(T) = c^\ast |T| \ .
\end{equation}

Then, it is easy to see that $w_k' = v_k'$, $w$ is additive and $N$ is an equivalence set in $w$.

\bigskip

(II) The classes of (strictly) convex, (strictly) weakly superadditive and (strict\-ly) monotone games: Let $M > \max_{T \subset N} | v_k' (T) |$, and for the coalitions on which $w$ is arbitrarily defined ($T \cap (S \cup \{k\}) = \emptyset$, $T \neq \emptyset$, see Lemma \ref{lemma3}): let

\begin{equation}\label{eq3}
w(T) = M |N| (|T|+1)^{|T|} \ .
\end{equation}

(A) Then it is easy to see that if $v$ is a (strictly) weakly superadditive, (strictly) monotone game, then so is $w$ (for the other properties see Lemma \ref{lemma3}). 

(B) Next we show that, if $v$ is a (strictly) convex game, then so is $w$. First notice that game $z \in \mathcal{G}^N$ is strictly convex, if and only if for each $i \in N$, $T,Z \subseteq N \setminus \{i\}$ such that $Z \subset T$: $z_i' (Z) < z_i' (T)$ (see \eqref{lemma50}). 

Let $l \in N \setminus (S \cup \{k\})$ and $T,Z \subseteq N \setminus \{l\}$ be such that $Z \subset T$, then

\begin{equation*}
\begin{split}
w_l' (T) =  w(T \cup \{l\}) - w(T) \\
=  w((T \cup \{l\})\setminus (S \cup \{k\})) + \sum \limits_{i=1}^m w_{l_i}' (((T \cup \{l\}) \setminus (S \cup \{k\})) \cup \{l_1,\ldots,l_{i-1}\}) \\ 
-  w(T \setminus (S \cup \{k\})) - \sum \limits_{i=1}^m w_{l_i}' ((T \setminus (S \cup \{k\})) \cup \{l_1,\ldots,l_{i-1}\}) \ ,
\end{split}
\end{equation*}

\noindent and

\begin{equation*}
\begin{split}
w_l' (Z) =  w(Z \cup \{l\}) - w(Z) \\
=  w((Z \cup \{l\})\setminus (S \cup \{k\})) + \sum \limits_{i=1}^n w_{l_i}' (((Z \cup \{l\}) \setminus (S \cup \{k\})) \cup \{l_1,\ldots,l_{i-1}\}) \\ 
-  w(Z \setminus (S \cup \{k\})) - \sum \limits_{i=1}^n w_{l_i}' ((Z \setminus (S \cup \{k\})) \cup \{l_1,\ldots,l_{i-1}\}) \ ,
\end{split}
\end{equation*}

\noindent where $m = |(S \cup \{k\}) \cap T|$, $n = | (S \cup \{k\}) \cap Z|$ and  $\{l_1,\ldots,l_n\} = (S \cup \{k\}) \cap Z = (S \cup \{k\}) \cap (Z \cup \{l\}) \subseteq (S \cup \{k\}) \cap (T \cup \{l\}) = (S \cup \{k\}) \cap T = \{l_1,\ldots,l_m\}$.

Notice that, if $T \setminus (S \cup \{k\}) = Z \setminus (S \cup \{k\})$, then the proof is complete. Therefore, w.l.o.g. we can assume that $Z \setminus (S \cup \{k\}) \subset T \setminus (S \cup \{k\})$. Game $v$ is a (strictly) convex game, $S \cup \{k\}$ is an equivalence set in $w$ and $n < |N|$ so

\begin{equation}\label{eqkell1}
\begin{split}
\sum \limits_{i=1}^m w_{l_i}' (((T \cup \{l\}) \setminus (S \cup \{k\})) \cup \{l_1,\ldots,l_{i-1}\}) \\ -\sum \limits_{i=1}^m w_{l_i}' ((T \setminus (S \cup \{k\})) \cup \{l_1,\ldots,l_{i-1}\}) \\ - \sum \limits_{i=1}^n w_{l_i}' (((Z \cup \{l\}) \setminus (S \cup \{k\})) \cup \{l_1,\ldots,l_{i-1}\}) \\ + \sum \limits_{i=1}^n w_{l_i}' ((Z \setminus (S \cup \{k\})) \cup \{l_1,\ldots,l_{i-1}\}) > -2M |N| \ .
\end{split}
\end{equation}

\noindent On the other hand, from \eqref{eq3}

\begin{equation}\label{eqkell2}
\begin{split}
w((T \cup \{l\})\setminus (S \cup \{k\})) - w(T \setminus (S \cup \{k\})) \\ - w((Z \cup \{l\})\setminus (S \cup \{k\})) + w(Z \setminus (S \cup \{k\})) \\
= M |N| \left( (|(T \cup \{l\})\setminus (S \cup \{k\})| + 1)^{|(T \cup \{l\})\setminus (S \cup \{k\})|} \right. \\
- \left. (|T \setminus (S \cup \{k\})| + 1)^{|T \setminus (S \cup \{k\})|} \right. \\
- \left. (|(Z \cup \{l\})\setminus (S \cup \{k\})| + 1)^{|(Z \cup \{l\})\setminus (S \cup \{k\})|} \right. \\
+ \left. (|Z \setminus (S \cup \{k\})| + 1)^{|Z \setminus (S \cup \{k\})|}\right) > 2 M |N| \ .
\end{split}
\end{equation}

\noindent Summing up \eqref{eqkell1} and \eqref{eqkell2} 

\begin{equation}\label{eq4}
w_l' (T)  - w_l' (Z) > 0 \ .
\end{equation}

\noindent Since $l \in N \setminus (S \cup \{k\})$ and $T,Z \subseteq N \setminus \{l\}$, $Z \subset T$ were arbitrarily chosen, $w$ is (strictly) convex (for the other properties see Lemma \ref{lemma3}).

\bigskip

(III) The class of (strictly) concave and (strictly) subadditive games.

(A) Notice that a game $v$ is (strictly), if and only if $\bar{v}$ is (strictly) convex (see \eqref{dualka}). Therefore, see Remark \ref{rem2}, from Point (II) the class of (strictly) concave games is an $M$-closed class of games.

(B) The class of (strictly) weakly subadditive games: It is worth noticing that the dual of a (strictly) subadditive or a (strictly) weakly subadditive game is not necessarily (strictly) superadditive or (strictly) weakly superadditive respectively, e.g. $v = (4,4,4,4,4,4,7)$ is strictly subadditive, but $\bar{v}$ is not weakly superadditive. Moreover, the dual of a (strictly) superadditive or a (strictly) weakly superadditive game is not necessarily (strictly) subadditive or (strictly) weakly subadditive, e.g. $v = (0,0,0,3,1,2,4)$ is strictly superadditive, but $\bar{v}$ is not weakly subadditive. 

Let $M =  v_k' (\emptyset)$, and for the coalitions on which $w$ is arbitrarily defined ($T \cap (S \cup \{k\}) = \emptyset$, $T \neq \emptyset$, see Lemma \ref{lemma3}): let

\begin{equation}\label{eq5}
w(T) = M |T|^2 \ .
\end{equation}

Then it is easy to see that, if $v$ is a (strictly) weakly subadditive game, then so is $w$ (for the other properties see Lemma \ref{lemma3}). 

\bigskip

Finally we can apply Theorem \ref{tetel1}.
\end{proof}

Notice that not all the classes of games (defined in the Preliminaries) are $M$-closed, the classes of essential, (strictly) superadditive, (strictly) subadditive games are not $M$-closed. The next example shows this fact.

\begin{example}\label{pl2}
(1) Let $v = (0,0,10,50,0,0,20)$, where $S = \{1,2\}$ is an equivalence set in $v$. Game $v$ is essential. However, the only game $w$ such that $N$ is an equivalence set in $w$, and $w_3' = v_3'$ is $w = (10,10,10,10,10,10,-20)$, but $w$ is not essential.

(2) Let $v = (0,0,0,10,51,51,51,51,51,51,62,62,62,62,103)$, where $S$ $= \{1,2,$ $3\}$ is an equivalence set in $v$. Game $v$ is strictly superadditive. However, the only game $w$ such that $N$ is an equivalence set in $w$, and $w_4' = v_4'$, is $w = (10,10,10,10,61,61,61,61,61,61,72,72,72,72,113)$, but $w$ is not superadditive. For the subadditive case take $-v$.
\end{example}

\begin{remark}
If $|N| \leq 3$, then the classes of (strictly) superadditive, (strictly) subadditive games coincide with the classes of (strictly) weakly superadditive, (strictly) weakly subadditive games respectively, hence they are $M$-closed. Furthermore, if $|N| = 2$, then the class of essential games coincides with the class of strictly superadditive games, hence it is $M$-closed.
\end{remark}

Although the above mentioned classes of games are not $M$-closed, so Theorem \ref{tetel1} cannot be applied to them, \cite{Young1985}'s axiomatization works for them (see p. 71 in \cite{Young1985}).

%\bibliographystyle{/home/pmiklos/Dropbox/Munka/Styles/ANOR/spbasic}
%\bibliography{/home/pmiklos/Dropbox/Munka/Hivatkozasok/referencesPMP.bib}

\end{document}